\documentclass[aps,pre,twocolumn,reprint,tightenlines,
amsmath,amssymb,amsfonts,showpacs,showkeys,floatfix]{revtex4-1}
\usepackage{graphicx}
\usepackage{dcolumn}
\usepackage{bm}
\usepackage{enumerate}
\usepackage{array}
\usepackage{multirow}
\usepackage{enumitem}
\usepackage{color}
\usepackage{braket}
\usepackage{csquotes}
\newcommand{\beq}{\begin{equation}}
\newcommand{\eeq}{\end{equation}}
\begin{document}
\title{Multipolar phase in frustrated spin-1/2 and 1 chains}
\author{Aslam Parvej}
\email{aslam12@bose.res.in}
\author{Manoranjan Kumar}
\email{manoranjan.kumar@bose.res.in}
\affiliation{S. N. Bose National Center for Basic Sciences, Kolkata,700098, India}
\date{\today}
\begin{abstract}
The $J_1-J_2$ spin chain model with nearest neighbor $J_1$ and next nearest neighbor 
anti-ferromagnetic $J_2$ interaction is one of the most popular frustrated magnetic models. 
This model system has been extensively studied theoretically and applied to explain the 
magnetic properties of the real low-dimensional materials. However, existence of different 
phases for the $J_1-J_2$ model in an axial magnetic field $h$ is either not understood 
or has been controversial. In this paper we show the existence of higher order $p>4$ 
multipolar phase near the critical point $(J_2/J_1)_c=-0.25$. The criterion to detect the 
quadrupolar or spin nematic (SN)/spin density wave of type two (SDW$_2$) phase using the 
inelastic neutron scattering (INS) experiment data is also discussed, and INS data of 
LiCuVO$_4$ compound is modelled. We discuss the dimerized and degenerate ground state 
in the quadrupolar phase. The major contribution of binding energy in the spin-1/2 
system comes from the longitudinal component of the nearest neighbor bonds. We also study 
spin nematic/SDW$_2$ phase in spin-1 system in large $J_2/J_1$ limit.  
\end{abstract}
\maketitle
\section{Introduction}
\label{Sec:I}
Interaction induced frustration and confinement of electrons in an one dimensional (1D) magnetic 
system generates many exotic phases \cite{Chubukov1991,Vekua2007,HikiharaMP2008,SudanLauchili}. Some 
of these phases can have well defined order parameters, whereas other phases can have hidden order 
parameter. The 1D spin-1/2 systems with an isotropic $J_1-J_2$ model \cite{MajumdarGhosh,TonegaHarada,
Chitra1995,WhiteAffleck,ManuScaling2007,ManuModfDMRG,Manu2010BOW, Sandvik, ManuDecouple,ManuAslamJPCM,
ItoiQuin2001,Haldane1982,HamadaFQCP,OkamotoNomura,Bursill,SirkerPRB,Mahdavifar} in the presence of an 
axial magnetic field have been extensively studied \cite{HikiharaMP2008,SudanLauchili, Dimitriev2006,
Meisner2006,Vekua2007,Kecke2007,MeisnerMcCulloch,Furukawa,SirkerPRB}. The $J_1-J_2$ model in an axial 
magnetic field $h$ is written as
\begin{equation}
H(J_1,J_2)=J_1\sum_{n}{S}_{n} \cdot {S}_{n+1}+J_2\sum_{n}{S}_{n} \cdot {S}_{n+2} -h\sum_{n} S_{n}^{z},
\label{eq1}
\end{equation}
where $J_1$ and $J_2$ are exchange interaction strengths between nearest neighbor (NN) and next nearest neighbor 
(NNN) spins, respectively. \\
\indent
The model with a ferromagnetic $J_1$ shows many interesting phases like spin liquid \cite{ManuAslamJPCM,
ItoiQuin2001,HamadaFQCP,SirkerPRB,Mahdavifar}, dimer \cite{ManuAslamJPCM,ItoiQuin2001,HamadaFQCP,SirkerPRB,Mahdavifar}, 
chiral vector \cite{Chubukov1991,AslamJMMM}, spin multipolar \cite{HikiharaMP2008,SudanLauchili}, decoupled phase 
\cite{ManuAslamJPCM}. The spin liquid phase is gapless and possesses quasi-long range order \cite{WhiteAffleck,ItoiQuin2001}. 
The dimer phase is gapped in nature, and the spin-spin correlation decays exponentially \cite{Manu2010BOW,WhiteAffleck,
ItoiQuin2001}. This model has been extensively used for modelling the magnetization properties of LiCuSbO$_{4}$ 
\cite{SianManuINS2012}, LiCu$_{2}$O$_{2}$ \cite{ParkMultifer}, (N$_2$H$_5$)CuCl$_3$ \cite{N2H5CuCl3}, 
Rb$_{2}$Cu$_{2}$Mo$_{3}$O$_{12}$ \cite{Rb2Cu2Mo3O12}, Li$_{2}$CuZrO$_{4}$ \cite{Li2CuZrO4}, Ba$_{3}$Cu$_{3}$In$_{4}$O$_{12}$, 
and Ba$_{3}$Cu$_{3}$Sc$_{4}$O$_{12}$ \cite{BaCuIn,BaCuSc}. In the chiral vector phase, both spin parity and inversion 
symmetry are spontaneously broken \cite{ManuSpinParity}. This phase has been studied extensively because of its potential 
application in improper multiferroic systems \cite{MathurMultifer,KatsuraMultifer}.\\
\indent
The field theoretical and numerical studies by Hikihara {\it{et al.}} suggest that metamagnetic or spin 
multipolar phase exist in the presence of the high axial magnetic field $h$ for ferromagnetic $J_1$ 
\cite{HikiharaMP2008}. These multipolar phases have hidden order parameters. In this model multipoles 
of order $p$ depend on the $J_2/J_1$ ratio \cite{HikiharaMP2008,SudanLauchili}, and the nomenclature of 
each phase is done based on the number of bound magnons in the systems i.e., the number of paired magnons 
$p$ in dipolar, quadrupolar, octupolar and hexadecapolar phases are $1,2,3,4$, respectively. The quadrupolar 
phase is a Tomonaga-Luttinger liquid of hard core bosons \cite{HikiharaMP2008}, and each boson is made up 
of two magnons. In this phase the correlations between bosons and density fluctuations follow a power law. 
However, the boson propagator is dominant over the density fluctuations in this phase \cite{HikiharaMP2008}. 
In his seminal work Chubukov predicts that this phase has dimerized ground state (gs) \cite{Chubukov1991}, but 
Hikihara {\it{et al.}} show the absence of dimerization \cite{HikiharaMP2008}. In the large $J_2/J_1$ regime, 
field theoretical calculations show that the SDW$_2$ phase exist in low magnetic field, whereas SN phase exists 
in the narrow range of magnetic field near the saturation field \cite{HikiharaMP2008}. The numerical 
calculations in $J_2/|J_1|=\alpha > 0.6$ show the finite binding energy of magnon even for a small field 
\cite{AslamJMMM}.\\
\indent
The order parameter of the SN phase $\langle S_i^{+} S_{j}^{+} \rangle$ is defined in ref. 
\cite{OlegBalents2014, Chubukov1991,Penc}. It is hidden in nature, although the probes like the 
INS \cite{EnderleDynamic,SianManuINS2012} and the resonant inelastic X-Ray scattering (RIXS) \cite{RIXS} 
methods can indirectly measure these phases. The nematic phase in LiCuVO$_4$ compound is confirmed by using 
the INS data of dynamical structure factor \cite{EnderleDynamic}, and NMR data of this compound shows a sharp 
single and solitary line which moves with magnetic field \cite{SatoNMR1,SatoNMR2}. In this paper we try to 
show that there is characteristic feature of INS measurement for the SDW$_2$ and SN phase.\\
\indent
In this model there are many unsettled issues such as, the metamagnetic phase in the small $J_2/J_1$ 
regime has been completely unexplored, and is difficult to characterize because of very small 
gaps. We have shown the gs degeneracies in the odd $S^z$ sectors \cite{AslamJMMM}, but dimer order 
parameter $B$ is vanishingly small in this sector. The existence of quadrupolar phase in 
spin-1 systems is controversial, as steps of two in magnetization-$h$ curve is absent \cite{Arlego2011,Kolezhuk2012}, 
whereas the other studies for general spin show the existence of this phase. We explore this phase for the 
spin-1 system using the Hamiltonian in Eq. \ref{eq1}.\\ 
\indent
The rest of the paper goes in the following sequence. In section II the numerical techniques and 
accuracy of results are discussed. Results are discussed in section III. We start with the higher 
order multipolar phase and the relation between the pitch angle $\theta$ and magnetization $M$. The 
quadrupolar phase is discussed thereafter. The dynamical properties in quadrupolar phase of spin-1/2 $J_1-J_2$ 
model are discussed in subsection B. The dynamical properties of LiCuVO$_4$ are also discussed in 
this subsection B. The dimer phase in the SN/SDW$_2$ phase is presented in the subsection 
C. The results for spin-1 for the same model are discussed in the section IV. The discussion of all 
the results is done in the next section V. 
\section{Numerical methods}
\label{Sec:II}
The Density matrix renormalization group method (DMRG) is a state of art numerical technique to 
calculate accurate gs and a few low lying excited energy states of strongly interacting quantum 
systems \cite{WhiteDMRG,KarenDMRG}. It is based on systematic truncation of irrelevant degrees of 
freedom. We use modified DMRG algorithm, where four new sites are added to avoid the multiple 
time of renormalization of operators in the superblock. The modified DMRG has better convergence 
and also has sparse Hamiltonian matrix of superblock for the model Hamiltonian in Eq. \ref{eq1}, 
compared to the conventional DMRG where only one site is added in each block at every step 
\cite{ManuModfDMRG}. The number of eigenvectors of the density matrix retained up to $m=400$ to 
maintain the truncation error of density matrix eigenvalues less than $10^{-10}$. In the worst case error 
in the energy is less than $0.01\%$. The DMRG is used for calculating various properties of large system 
sizes up to $N = 368$ chain with open boundary condition (OBC). The number of finite DMRG sweeps 
required for an accurate gs and spin correlation function in the different $S^z$ sectors is 
approximately 20. Recently developed PBC algorithm is also employed for calculating the accurate 
gs and the correlation functions \cite{DayaPBC}. The dynamical structure factor is calculated using 
the correction vector method \cite{RamseshaCorrection,JeckelmannCorrection,OnishiJapan}.
\section{Results}
\label{Sec:III}
The quantum phase diagram of $J_1-J_2$ model in an axial magnetic field given in Eq. (\ref{eq1}) 
consists of numerous phases such as the vector chiral (VC) \cite{Chubukov1991,AslamJMMM}, the dimer 
\cite{MajumdarGhosh,TonegaHarada,Chitra1995,WhiteAffleck,ManuScaling2007,ManuModfDMRG,Manu2010BOW,
ManuDecouple,ManuAslamJPCM,ItoiQuin2001,Haldane1982,HamadaFQCP,OkamotoNomura,Bursill,SirkerPRB,Mahdavifar}, 
the decoupled chain \cite{ManuDecouple,ManuAslamJPCM}, and multipolar/SDW$_n$ phases \cite{HikiharaMP2008,SudanLauchili}. 
In this paper, the SN/SDW$_2$ phase and other higher order multipolar phases are discussed. This section is divided into 
three subsections. In subsection A, multipolar phases for spin-1/2 are discussed in the beginning; SN/SDW$_2$ 
phase is presented in later part of subsection A. The general observations about dynamical property and $M-h$ curve 
in quadrupolar phase are presented. We model the dynamical structure factor of LiCuVO$_4$ and also compare our results 
with the experimental data available in literature \cite{EnderleDynamic,MourigalFieldMom} in subsection B. The dimer in 
SN/SDW$_2$ phase is presented in subsection C. 
\subsection{Multipolar phases in $S=1/2$}
\label{Sec:A}
\begin{figure}[t]
\begin{center}
\includegraphics[width=0.5\textwidth]{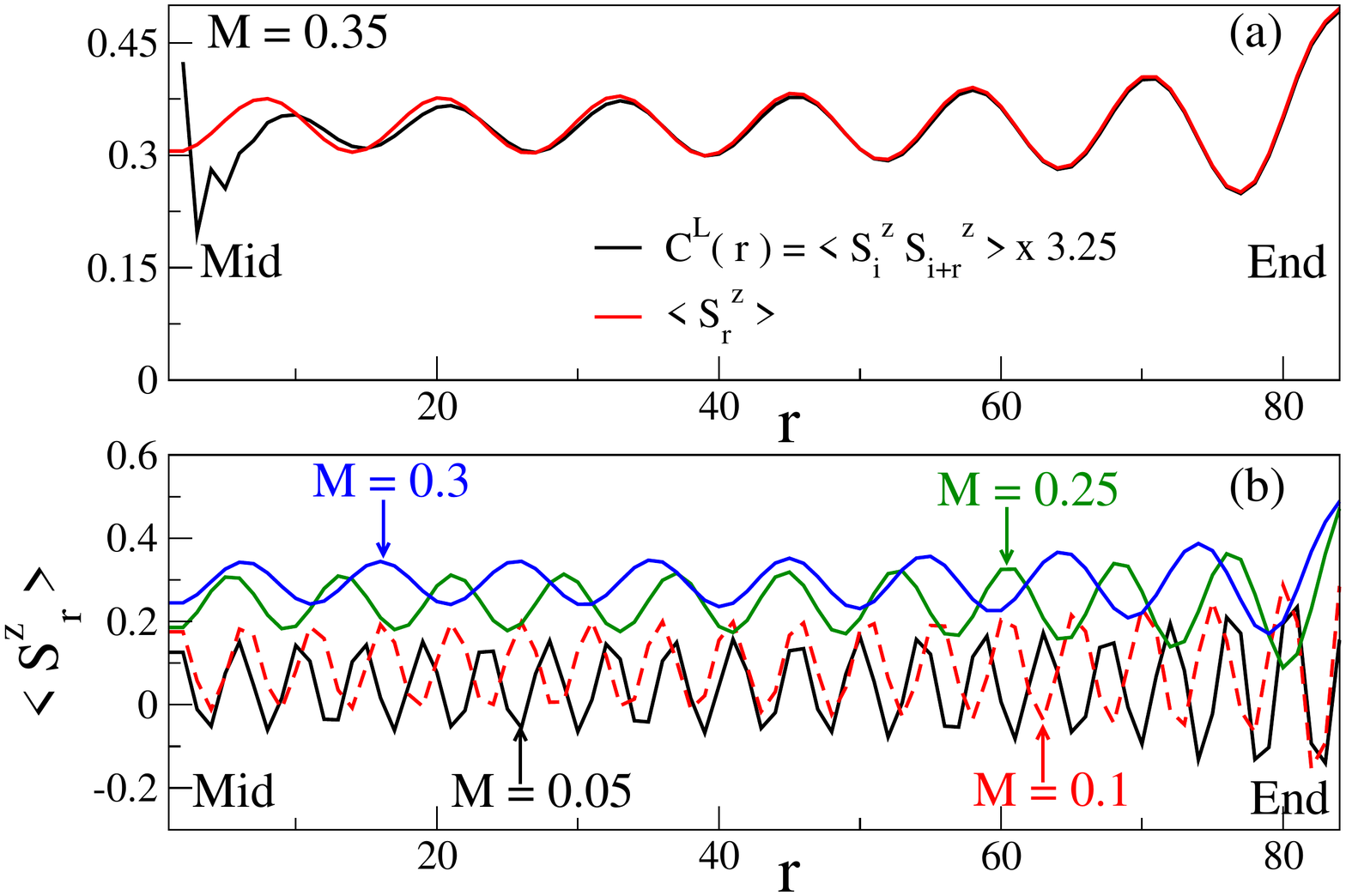}
\caption{The upper panel (a) shows the spin density $\langle S_r^z\rangle$ and longitudinal correlation function 
$C^L(r)$ for $M=0.35$. $C^L(r)$ is multiplied by 3.25 time and x-axis of this plot is shifted by 1 to match the 
magnitude and phase of $\langle S_r^z\rangle$. In the lower panel, spin densities $\langle S^z_r\rangle$ for 
$M=0.05,0.1,0.25$ and 0.3 are shown. For the $C^L(r)$ mid site of the chain is kept as reference site.}
\label{fig1}
\end{center}
\end{figure}

\begin{figure}[b]
\begin{center}
\includegraphics[width=0.5\textwidth]{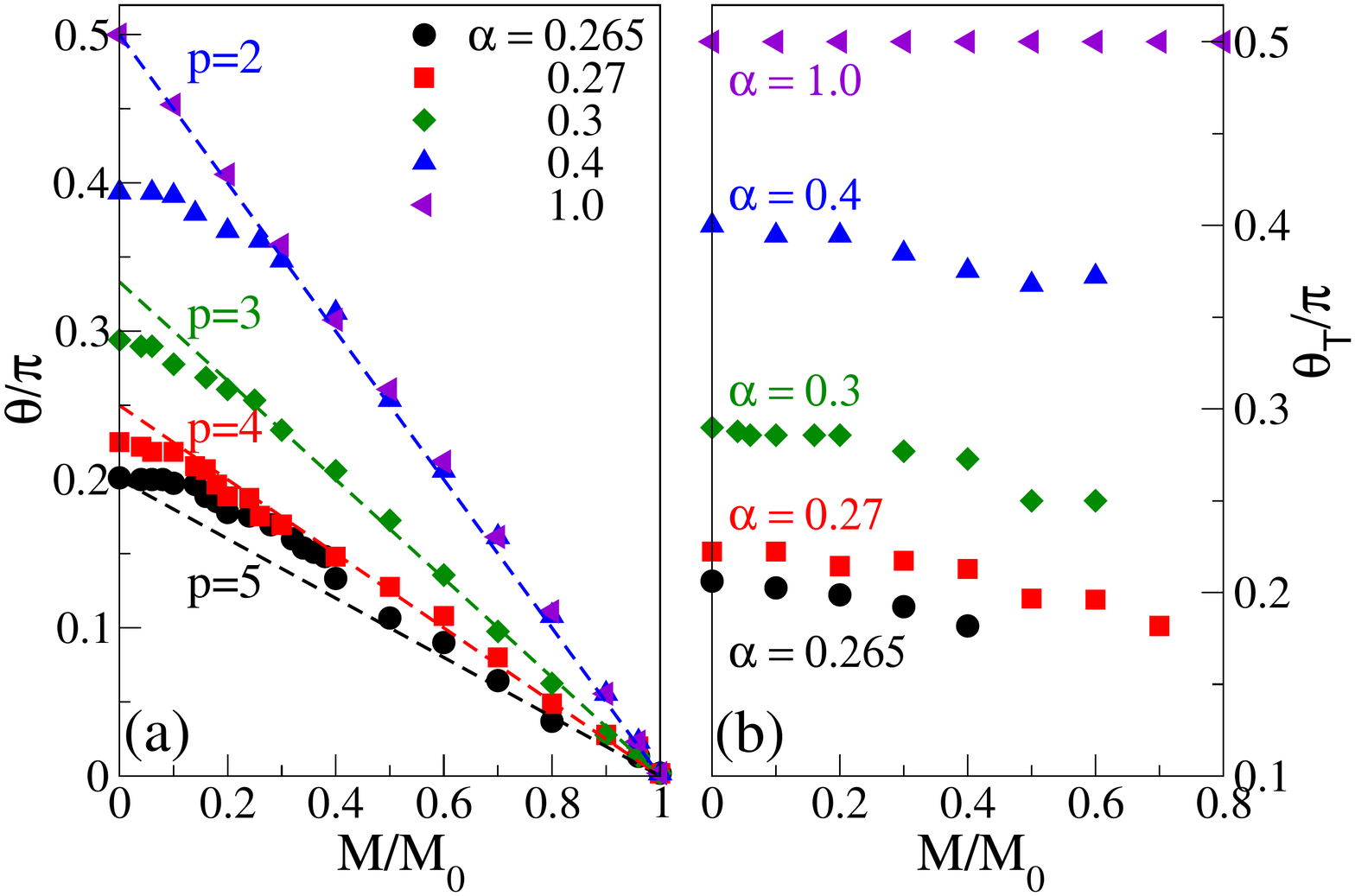}
\caption{In panel (a); Pitch angle $\theta$ is calculated from $\langle S_r^z\rangle$ and $C(r)$. In panel (b) 
transverse pitch angle $\theta_T$ is calculated from transverse correlation as a function of $M/M_{0}$ for 
$\alpha= 0.265,0.27,0.3,0.4$ and 1.0 are shown. The dashed lines in the left panel are fitted lines with the 
equation $\frac{\theta}{\pi} = \frac{1}{p}(1-\frac{M}{M_0})$ where $p$ is the order of the multipolar phase.}
\label{fig2}
\end{center}
\end{figure}

The multipolar phase and the spin density wave in the $J_1-J_2$ model for spin-1/2 chain in the presence of magnetic 
field $h$ are discussed in this part. We notice that there is a level crossing from ferromagnetic to singlet gs at 
$\alpha_{c}=0.25$ \cite{HamadaFQCP}, and near to the critical point $\alpha_c$, but $\alpha > 0.25$ limit, multiple 
magnons bind to form multipoles below the saturation magnetic field. It is also noted that number of $p$ changes 
rapidly with $\alpha$. In this paper, multipolar phase with order $p$ is explored based on the magnetic steps, 
pitch angle $\theta$ of spin density, and spin correlations in the gs at a finite magnetic field $h$. The angle 
between two nearest neighbor spin is called pitch angle $\theta$ and is defined as
\begin{equation}
\begin{split}
\theta=\frac{2 \pi}{L}, 
\label{eq1p}
\end{split}
\end{equation}
where $L$ is the smallest distance between spins whose pitch angle differs by $2\pi$. The field theoretical 
bosonization calculations \cite{HamadaFQCP} suggest that for $\alpha > 0.25$, the system shows SDW$_n$ in low 
magnetic field, whereas it shows multipolar phase at high magnetic field \cite{HikiharaMP2008}. The multipolar 
correlation of order $p$ or boson propagator 
$\langle S_i^{+} S^{+}_{i+1} \cdot \cdot \cdot S^{+}_{p+i-1} S^{-}_{i+r+1} S^{-}_{i+r+2} 
\cdot \cdot \cdot S^{-}_{i+r+p-1} \rangle$ is written \cite{HikiharaMP2008} as 
\begin{equation}
\begin{aligned}
\langle S^{+}_{0} \cdot \cdot \cdot S^{+}_{p-1} S^{-}_{r} \cdot \cdot \cdot S^{-}_{r+p-1} \rangle 
=(-1)^{r}\langle b_{0}b_{r}^{\dagger}\rangle \\
=\frac{A_{m}(-1)^{r}}{|r|^{1/\eta}}-\frac{\widetilde{A}_{m}(-1)^{r}}{|r|^{\eta+1/\eta}} \cos(2\pi\rho r)
+ \cdot \cdot \cdot,
\label{eq2}
\end{aligned}
\end{equation}
where $A_m$ and $\widetilde{A}_m$ are constants, $\eta$ is twice of the Luttinger liquid parameter, and 
$r$ represents distance. The density-density correlation is written as
\begin{equation}
\begin{split}
C^L(r)=\langle S^{z}_{0} S^{z}_{r} \rangle & =\Bigg \langle \bigg (\frac{1}{2}-pb_{0}^{\dagger}b_{0} 
\bigg) \bigg (\frac{1}{2}-pb_{r}^{\dagger}b_{r} \bigg) \Bigg \rangle \\
& = M^{2}-\frac{p^{2}\eta}{4\pi^{2}r^{2}}+\frac{A_{z} \cos(2\pi\rho r)}{|r|^{\eta}}+ \cdot \cdot \cdot,
\label{eq3}
\end{split}
\end{equation}
where $\rho=\frac{1}{p}(1-\frac{M}{M_0})$, $M_0$ is the saturation magnetization. The pitch angle 
$\theta=2\pi \rho$ varies with the magnetic field. The spin density $\langle S_{r}^{z} \rangle$ 
calculated from the field theoretical method is written \cite{HikiharaMP2008} as
\begin{equation}
\begin{aligned}
\langle S^{z}_{r} \rangle & = \frac{1}{2}(1-p)-pz(r;q); \\
z(r;q) & = \frac{q}{2\pi}-a\frac{(-1)^{r} \sin(qr)}{f_{\eta/2}(2r)}; \\
q & = \frac{2\pi N}{N+1}(\rho-\frac{1}{2});\\ 
f_{\nu}(x) & = \Bigg [\frac{2(N+1)}{\pi} \sin \bigg(\frac{\pi|x|}{2(N+1)}\bigg) \Bigg ]^{\nu}.
\end{aligned}
\label{eq4}
\end{equation}
\indent
The pitch angle $\theta_T$ in transverse direction can also be extracted from the transverse correlation 
function $C^T(r)=\langle(S^{x}_{i}S^{x}_{i+r}+S^{y}_{i}S^{y}_{i+r})\rangle$. However, the pitch angle 
$\theta$ in the longitudinal direction is calculated from $C^L(r)$ and spin density $\langle S^{z}_{r} \rangle$.\\
\indent
The $\langle S^{z}_{r} \rangle$ and $C^L(r)$ are shown in Fig. \ref{fig1} (a) for $M=0.35, \alpha=1.0$ and 
$N=168$. The $C^L(r)$ is scaled by 3.25 and $r$ is shifted by 1.0 unit to match the magnitude of 
$\langle S^{z}_{r} \rangle$. Interestingly, the complex looking equation of $\langle S^{z}_{r} \rangle$ in 
Eq. (\ref{eq4}) has similar variation as that of $C^L(r)$. All the $\langle S^{z}_{r} \rangle$ and $C^L(r)$ 
give the same pitch angle. The Friedel oscillation at the edge of chain is seen in both $\langle S^{z}_{r} \rangle$ 
and $C^L(r)$. The spin densities are plotted in Fig. \ref{fig1} (b) for $M =0.05,0.1,0.25$ and $0.3$. The amplitude 
of $\langle S^{z}_{0} S^{z}_{r} \rangle$  decreases with the distance, whereas $|\langle S^{z}_{r} \rangle|$ at site 
$r$ is more or less constant with $r$. Therefore, it is easier to calculate $\theta$ from $\langle S^{z}_{r} \rangle$ 
than from $\langle S^{z}_{0} S^{z}_{r} \rangle$. We find that $\theta$ decreases with $M$ and reduces to zero at 
$M = 0.5$ for spin-1/2 system. These results are consistent with the Sudan {\it{et al.}} exact diagonalization 
results \cite{SudanLauchili}.\\
\indent
As shown in the Fig. \ref{fig1} (a) $\theta$ calculated from $\langle S^{z}_{r} \rangle$ and $C^L(r)$ are same, 
and it follows a linear relation with $M$. With $M$ the variations in $\theta_T$ of the transverse correlation 
functions $C^T(r)$ is less than 5$\%$. The accurate calculation of $\theta$ near $M_0 \approx 0.5$ requires larger 
system size, and for these calculations we have used $N=168$ for low magnetization and 368 for 
higher magnetization. The $\theta$ and $\theta_T$ are calculated from  the $\langle S^{z}_{r} \rangle$ and the 
$C^T(r)$, respectively, for $\alpha=0.265, 0.27, 0.3, 0.4,$ and 1.0 as a function of $M/M_{0}$ shown in 
Fig. \ref{fig2} (a) and 2 (b). The filled symbols are the DMRG calculations for $N=168$ with OBC, and dotted 
lines are fitted line with $\frac{\theta}{\pi} = \frac{1}{p}(1-\frac{M}{M_0})$ where $p$ is the order of the multipole.\\
In Fig. \ref{fig2} (a) we notice that the variation of the $\theta$ with $M$ shows linear relation 
$\frac{\theta}{\pi} = \frac{1}{p}(1-\frac{M}{M_0})$ especially at large $M$. For $ \alpha > 0.4 $ and 
large $M$, $\theta$ varies linearly with $M/M_{0}$ with a slope $2/p=1$. The linear behavior of $\theta$ deviates 
from straight line at low $M/M_{0}$ for $\alpha \leqslant 0.6$. The deviation point for 
$\alpha = 0.4$ is at $M/M_0 \approx 0.28$. In the VC phase $\theta$ depends weakly on $M$ as shown in 
Fig. \ref{fig2} (a). The phase boundary of the quadrupolar and the VC phase is estimated using the level 
crossing or magnetic step criterion as in ref. \cite{HikiharaMP2008,AslamJMMM}. For $\alpha \geq 0.4$  
results will be discussed in the later part of this section.\\
\indent
The three magnon bound phase or the triatic/SDW$_3$ phase occurs in the vicinity of $\alpha=0.3$ and  
$\frac{\theta}{\pi}$ is less than $ 0.26$ at $M < 0.21$ as shown in Fig \ref{fig2} (a). At large 
$M$ the slope of green line in Fig. \ref{fig2} (a) is $1/p=1/3$. The phase boundary of the triatic/SDW$_3$ and the VC 
phase can also be estimated from the deviation of the $\frac{\theta}{\pi}$ from linear relation as 
shown in Fig. \ref{fig2} (a). In fact $\theta$ weakly depends on $M$ in the VC phase and remains constant 
for the given value of $\alpha$, whereas it varies linearly with slope $1/p=1/3$ in the triatic/SDW$_3$ phase. 
The phase boundary of the triatic/SDW$_3$ and the VC phase calculated with this method is consistent with 
other calculations \cite{SudanLauchili,HikiharaMP2008}. The maximum value of $\theta$ for a multipole of 
order $p$ for a given $\alpha$ is $\pi/p$, and it decreases with the number of magnons or $p$. Our DMRG result 
shows that for $\alpha=0.265$ at large $M$, $p=5$ state shows up for $M/M_0 > 0.4$, and system show the $p=4$ 
state for intermediate magnetization $0.12 < M/M_0 <0.4$. The vector chiral phase sets in below the $M/M_0 \leq 0.12$. 
For $\alpha< 0.265$, $\theta$ calculations become difficult with the approximate numerical technique. In this 
limit energy states are closely spaced, and accurate determination of wavefunctions of the closely spaced  
energy levels is difficult.\\
\begin{figure}[b]
\begin{center}
\includegraphics[width=0.5\textwidth]{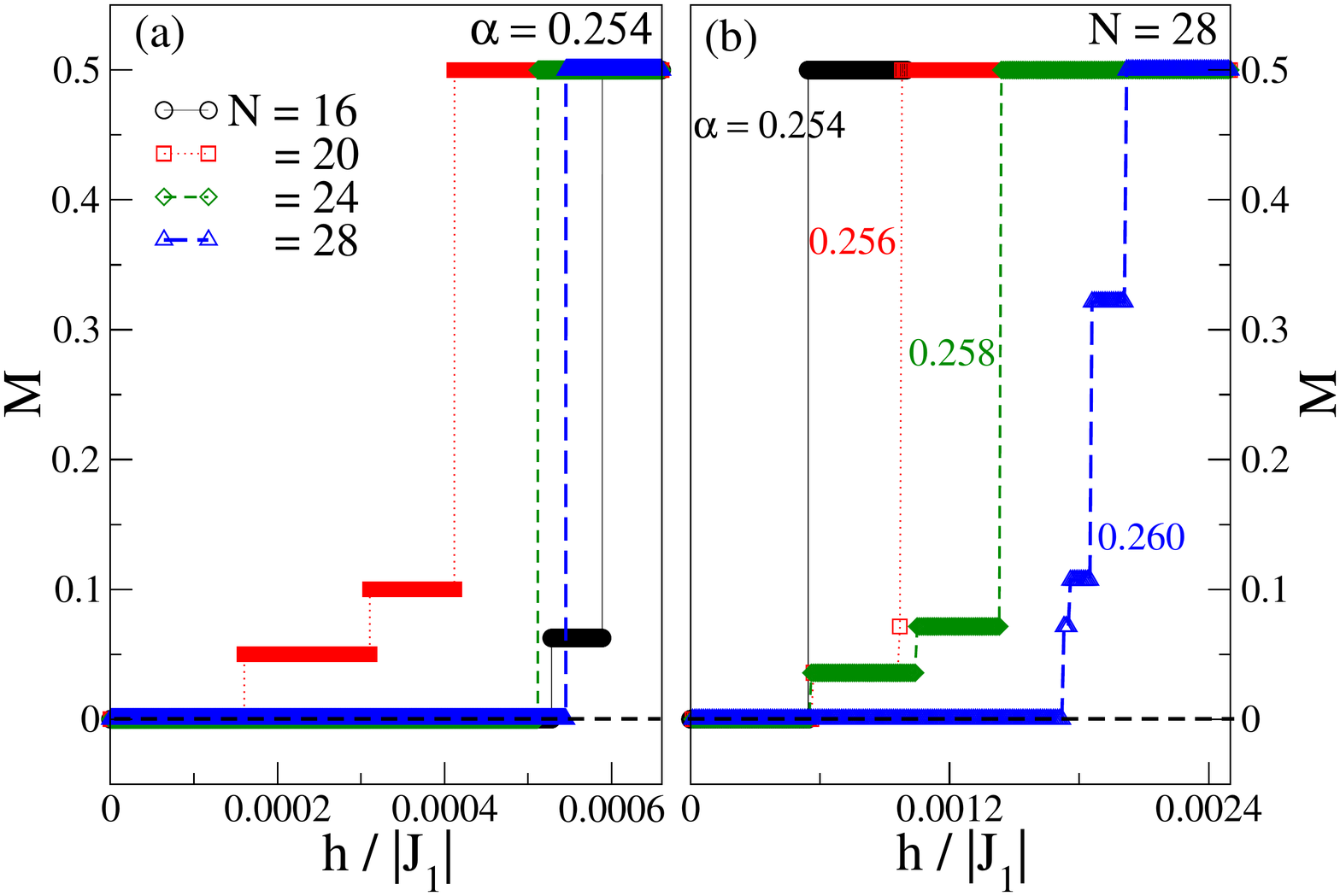}
\caption{(a) Magnetization vs. axial magnetic field $h$ for $\alpha=0.254$ is shown. The calculations are done for 
system sizes $N=16,20,24$ and 28 with PBC. (b) The $M$ vs. $h$ for $N=28$ for $\alpha=0.254,0.256,0.258$ and 0.26 
with PBC is shown.}
\label{fig3}
\end{center}
\end{figure}
\indent
In the quadrupolar phase the binding energy $E_b$ of the magnons defined in Eq. \ref{eq5} below is 
an important quantity to understand the condensation phenomenon. The $M-h$ curve is analyzed to see the effect of 
condensation of magnons. Near the critical point $\alpha_c=0.25$, energy level spacings are tiny. Therefore, 
to maintain the accuracy of results, the ED method is used to solve the Hamiltonian for systems with $N=16,20,24$ and 28. 
In Fig. \ref{fig3} (a), the finite size effect on the $M-h$ curve is shown for $\alpha=0.254$. The $N=16$ shows 
jumps of 1 and 7, whereas $N=20$ shows steps in $M$ of size 1 and 8. The gaps between the energy 
levels decrease with system size $N$, and for $N=24$ and 28 system shows steps of $N/2$. For $\alpha=0.254$ the 
value of $p$ can be equal or higher than $N/2$. The finite system size effect on the gaps is weak in this parameter 
regime. The $M-h$ curve for different values of $\alpha=0.254, 0,256, 0.258$ and 0.26 are  shown in Fig. \ref{fig3} (b) 
for $N=28$. We notice that the $h$ required for saturation increases with $\alpha$. The step size depends on 
$\alpha$ for example, at $\alpha=0.256$ and 0.258 system shows jumps of 1 and 12, whereas for $\alpha=0.26$ 
the jumps are 1, 2 and 6. The VC phase exists in the low $M$ limit, and the phase boundary decrease with $\alpha$. 
The magnetic steps or the order of multipole $p$ increases rapidly with  $1/\alpha$ near the critical point $0.25$, 
and the magnetic gaps decrease with $\alpha$ as shown in Fig. \ref{fig3} (a). Unfortunately we 
need large system size to confirm the large $p>14$, but these results are consistent with the prediction of 
existence of larger $p$ in ref. \cite{OlegBalents2016}.\\
\indent
In case of incommensurate spin density wave there are level crossings or the gs degeneracies, and these two 
degenerate states have opposite inversion symmetry \cite{AslamJMMM}. Our ED calculations show that the gs 
energies are degenerate at large magnetization for $\alpha < 0.4$ for both odd and even $S^z$ sectors. We 
calculate the z-component of VC order parameter 
$\kappa_{i}^{z}=\frac{1}{N}\sum_{i}\bra{\psi_{+}}(S_{i}^{+}S_{i+1}^{-}-S_{i}^{-}S_{i+1}^{+})\ket{\psi_{-}}$  
\cite{AslamJMMM}. In the multipolar phase, the $\kappa_{i}^{z}$ at large $S^z$ limit is non-zero for 
$0.25 < \alpha < 0.55$ and system sizes up to $N=28$. The $\kappa_{i}^{z}$ for $N=24$ system size is shown 
for different $M$ for $0.25 < \alpha < 1.0$ in ref. \cite{AslamJMMM}. \\
\indent
In the large $\alpha$ limit, the SDW$_2$ and SN phase exist in the presence of the magnetic field $h$. In the SN 
phase two magnons can condense to form a single boson \cite{HikiharaMP2008,AslamJMMM}, and this phase is determined 
based on the presence of magnetic step of two $(\Delta S^{z}=2)$ in $M-h$ curve \cite{HikiharaMP2008,AslamJMMM}, 
order parameter and various correlation functions. The order parameter for the quadrupolar phase is 
defined as in ref \cite{OlegBalents2014,Penc,Chubukov1991} 
\begin{equation}
\begin{aligned}
\rho_q=\bra{\psi_{n+2}} S_i^{+}S_j^{+}\ket{\psi_{n}},
\end{aligned}
\label{eq4p}
\end{equation}
where $\ket{\psi_{n}}$ and $\ket{\psi_{n+2}}$ are gs of $S^z=n$ and $n+2$ spin sector, but both of 
these are degenerate in the presence of an applied magnetic field.\\  
\indent
In this phase the variation of $\theta$, magnitude of $E_b$, $\rho_q$ and the dynamical structure factor 
$S(q,\omega)$ as a function of $M$ are calculated in the presence of magnetic field. The variation of 
magnetization $M$ with $h$ at $\alpha = 0.6$ is shown in Fig. \ref{fig4} (a) for chains with $N=104$ and 
168. The magnetic step of $\Delta S^z=2$ exists in the full range of $M$. The existing literature shows 
the SDW$_2$ phase at low magnetic field and SN type at high magnetic field 
\cite{HikiharaMP2008,OlegBalents2014,OlegBalents2016}. To analyze the quadrupolar phase, $\theta$ is plotted 
as a function of $M/M_{0}$ in Fig. \ref{fig4} (b) for two different $N=104$ and 168 chain. The dashed line 
indicate $\frac{\theta}{\pi} = \frac{1}{p}(1-\frac{M}{M_0})$ line with $p=2$. These calculations demonstrate 
weak size dependence of the pitch angle $\theta$. \\
\begin{figure}[t]
\begin{center}
\includegraphics[width=0.5\textwidth]{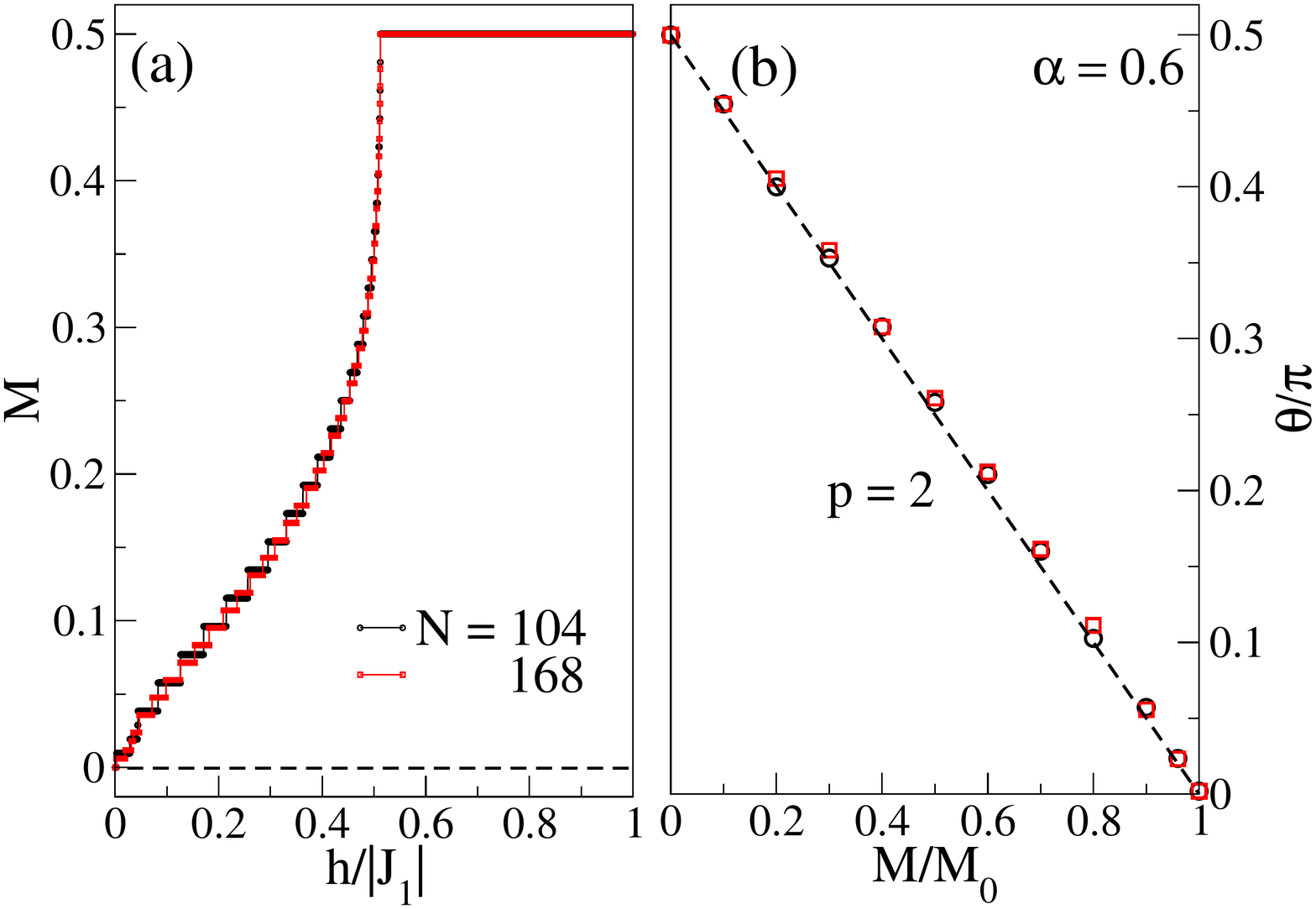}
\caption{(a) $M-h$ plot for $\alpha=0.6$ of system sizes $N=104$ and 168 using DMRG with OBC is shown. 
(b) The pitch angle $\theta$ as a function of magnetization $(M/M_{0})$ for system sizes $N=104$ and 
168 at $\alpha=0.6$ is shown. The dotted line is fitted line with $\frac{\theta}{\pi} = \frac{1}{p}(1-\frac{M}{M_0})$ 
where $p=2$.}
\label{fig4}
\end{center}
\end{figure}
\indent
The average binding energy of two magnons is defined as 
\begin{equation}
E_b(n)=\frac{1}{2}\Big [E(n+2)+E(n)-2E(n+1)\Big],
\label{eq5}
\end{equation}
where $E(n)$ is the energy of the system with even number of magnons $n$. The binding energy of two magnons 
in the SN/SDW$_2$ phase is shown as a function of $N$ in the inset of Fig. \ref{fig5}. The $|E_b|$ has weak 
finite size dependence in large $M$ limit, whereas it shows significant change with system size in low field 
limit. The finite size scalings are done for $M=0.05,0.1$ and $M=0.4$ at $\alpha=1.0$ for $N$ 
up to 200. $|E_b|$ increases with $M$ and has finite extrapolated value for $M>0.1$. However, $|E_b|$ at low 
magnetization $M=0.05$ is vanishingly small.\\
\indent
In Fig. \ref{fig5} the extrapolated values of $|E_b|$ as a function of $\alpha$ for different 
$M=0.05, 0.1, 0.15, 0.2, 0.25$ and 0.45 are shown. The error bars reflect the error in extrapolation 
and inaccuracy in DMRG calculations. We notice that $|E_b|$ increases with $\alpha$ and it attains a 
maximum value  around $\alpha_{m}(M)$ for a given $M$, and decreases thereafter. The value of $\alpha_{m}(M)$ 
increases with $M$. The $|E_b|$ increases with $M$ initially and either it saturates or decreases near 
the saturation magnetic field. This trend of $|E_b|$ is consistent with the calculations done by 
Onishi \cite{OnishiJapan}. \\
\indent
\begin{figure}[t]
\begin{center}
\includegraphics[width=0.5\textwidth]{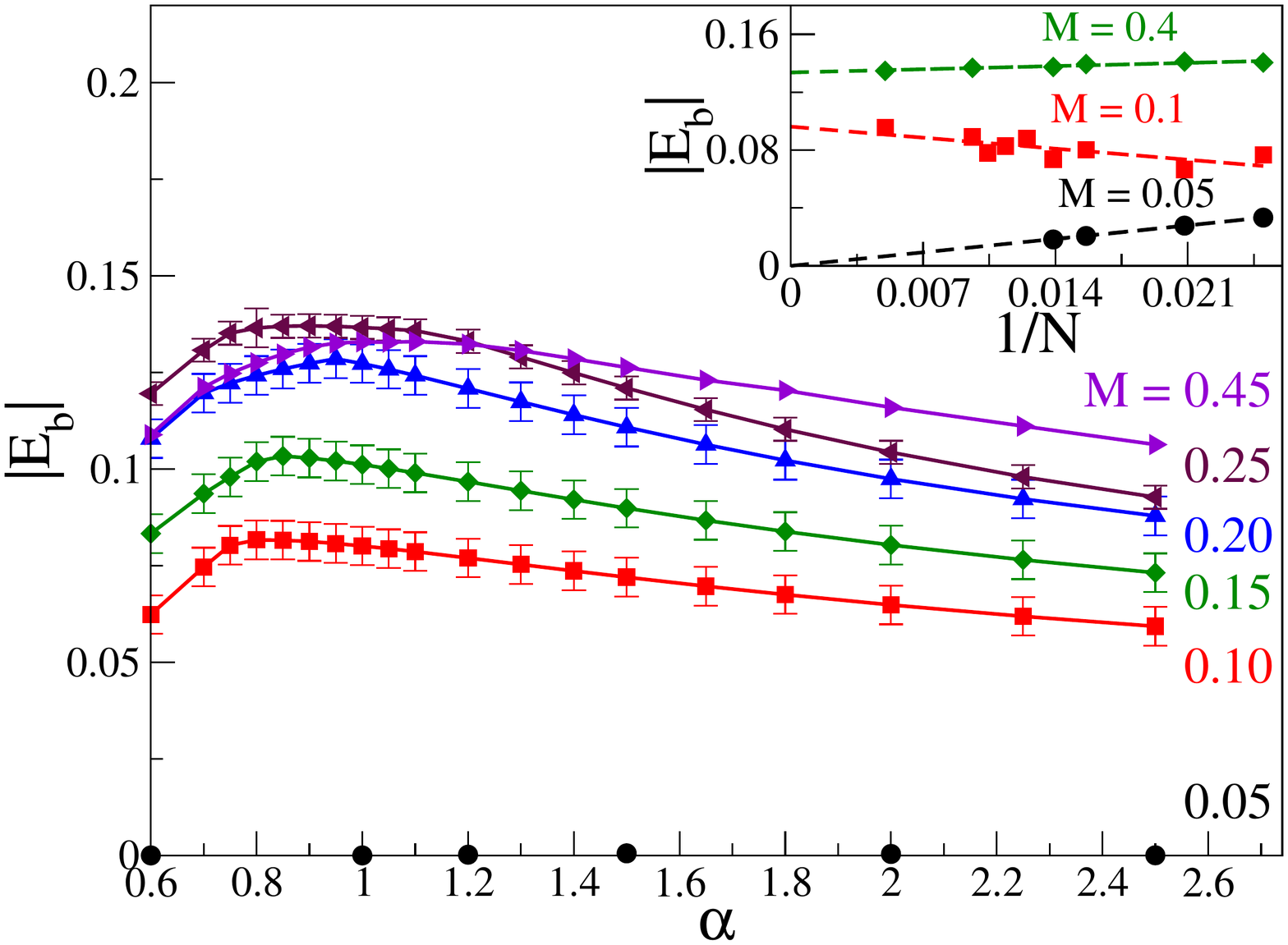}
\caption{The main figure shows the binding energy $|E_b|$ as a function of $\alpha$ for magnetization $M=0.05,0.1,
0.15,0.2,0.25$ and 0.45. In the Inset $|E_b|$ vs. $1/N$ for magnetization $M=0.05,0.1$ and 0.4 at $\alpha=1.0$ 
are shown.}
\label{fig5}
\end{center}
\end{figure}
The bond energies are analyzed to understand the contribution of different bonds in the $E_b$. In the large 
$\alpha$ limit the $J_1-J_2$ model for a chain behaves like a zigzag chain, and the next nearest neighbor 
interaction $J_2$ of the model act as interaction between the spins along the leg, whereas the nearest 
neighbor interaction $J_1$ becomes the interaction along the rung \cite{ManuModfDMRG}. The contribution of 
different bonds in the $E_b$ are calculated for $\alpha=1.0$ and at $M=0.25$ and 0.4 for a chain of sizes 
$N=$16, 20, 24 and 28 with PBC. However, data are shown only for $N=24$ and 28 in the table \ref{tb1}. The 
binding energy contribution of different bonds $E^{x,y}_{b}$ where $x$ stands for longitudinal $(L)$ or 
transverse $(T)$ and $y$ stands for leg $(L)$ or rung $(R)$. The $E_b$ is defined in terms of $E_b^{xy}$ as,
\begin{equation}
E_{b}(n)=\frac{1}{2}\Big [E^{x,y}_{b}(n+2)+ E^{x,y}_{b}(n)-2E^{x,y}_{b}(n+1) \Big].
\label{eq6}
\end{equation}

\begin{table}[b]
\centering
\caption{The contribution of different bonds in the $E_b$ are calculated for $\alpha=1.0$ and at $M=0.25$
and 0.4 for a chain of sizes $N=$16, 20, 24 and 28 with PBC}
\begin{tabular}{|c|c|c|c|c|c|c|}
\hline
\multirow{2}{*}{$\alpha$} & \multirow{2}{*}{$System$} & \multirow{2}{*}{$E_b$} & \multicolumn{2}{c|}{M = 0.25} & 
\multicolumn{2}{c|}{M = 0.4} \\ \cline{4-7}
& & & $N(24)$ & $N(28)$ & $N(24)$ & $N(28)$ \\
\hline
\multirow{8}{*}{1.0} & \multirow{3}{*}{Leg} & $E_{b}^{L,L}$ & -0.0408 & -0.0713 & -0.0332 & -0.0237 \\
& & $E_{b}^{T,L}$ &  0.1935 &  0.2208 & -0.2217 & -0.0815 \\  \cline{3-7}
& &      Total    &  0.1526 &  0.1495 & -0.2549 & -0.1051 \\  \cline{2-7}
& \multirow{3}{*}{Rung} & $E_{b}^{L,R}$  & -0.7103 & -0.6600 & -0.7150 & -0.7493 \\
& & $E_{b}^{T,R}$ &  0.2070 &  0.1963 &  0.6597 &  0.5756 \\ \cline{3-7}
& &      Total    & -0.5033 & -0.4637 & -0.0552 & -0.1737 \\ \cline{2-7}
& \multicolumn{2}{c|}{Total binding energy} & -0.3507 & -0.3142 & -0.3102 & -0.2788 \\ [1ex]
\hline
\end{tabular}
\label{tb1}
\end{table}
\indent
For $M=0.25$ the major contribution to $E_b$ are transverse component $E^{T,R}_b$, $E^{T,L}_b$ and $E^{L,R}$ 
,however, transverse component weakens the $E_b$ as shown in table \ref{tb1}. The $E^{L,L}_b$ decreases with 
system size, whereas $E^{T,L}_b$ increases with the system size. The magnitude of $E^{T,R}_{b}$ is significantly 
smaller than the $E^{L,R}_{b}$. The major contribution of $E_b$ comes from the $E^{L,R}_b$. The magnitude of 
$E^{T,R}_b$ is almost 1/3 of the $E^{L,R}_b$, but these two have opposite signs. However, both these quantities 
increase with $M$. For $M=0.4$ both along leg and rung transverse bonds contributions weaken the total $E_b$. 
The $E^{T,L}_b$ also decreases, whereas $E^{T,R}_b$ increases. The magnitude of $E^{L,R}_b$ and $E^{T,R}_b$ are 
very similar, but opposite to each-other for $M=0.4$. In conclusion rung contributes most of the $E_b$ in small 
$M$, but contribution of leg increases with $M$ of $M<M_0$. The $E_b$ is still small, however, $E^{T,L}_b$ is 
significantly large. \\
\begin{figure}[t]
\begin{center}
\includegraphics[width=0.5\textwidth]{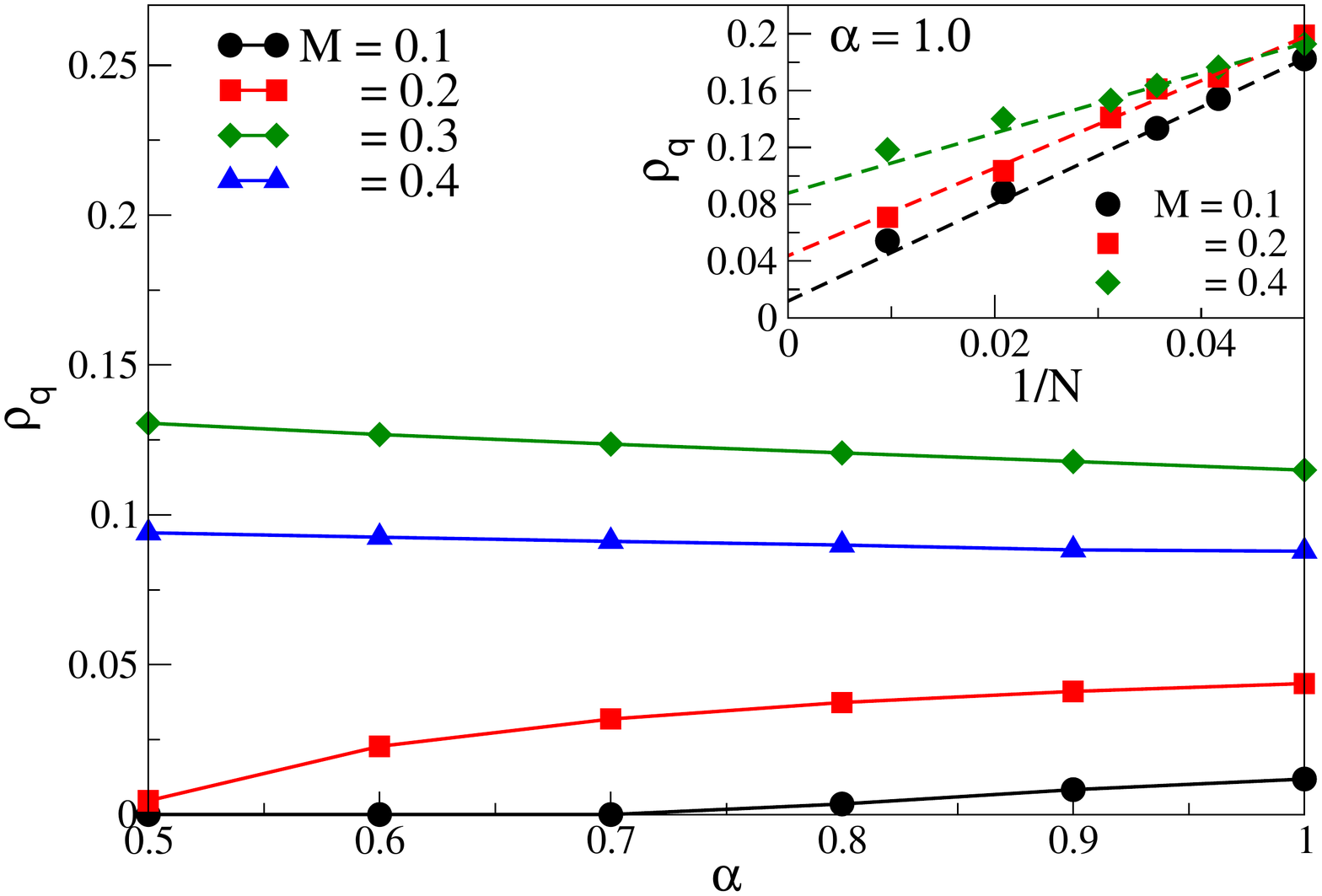}
\caption{The SN order parameters $\rho_q$ as a function of $\alpha$ for gs of $M=0.1$, 
0.2, 0.3, and 0.4 are shown in the main figure. The inset shows extrapolation of the $\rho_q$ 
as a function $1/N$ for magnetization $M=0.1,0.2$ and 0.4.}
\label{fig6}
\end{center}
\end{figure}
\indent
The quadrupolar phase is directly quantified in terms of the order parameter $\rho_q$ 
defined in Eq. \ref{eq4p}. In the inset of Fig. \ref{fig6} $\rho_q$ is extrapolated as a function of $1/N$ 
for $M=0.1$, 0.2 and 0.4. All the curves of $\rho_q$ follows the linear behavior with $1/N$. The extrapolation 
of $\rho_q$ is done with  system size upto $N=104$. The extrapolated values of $\rho_q$ are shown in the main 
Fig. \ref{fig6} for $M =0.1$, 0.2, 0.3, 0.4 for $ 0.5 \geq \alpha \geq 1.0$. The value of the $\rho_q$ is 
within the error limit for $M \leq 0.1$ for smaller $\alpha$. The $\rho_q$ increase with $M$, and varies 
slowly with $\alpha$. The  behavior of $\rho_q$ is quite consistent with $E_b$ as  both of these quantity 
increase with $M$. 
\subsection{Dynamical structure factor}
\label{Sec:B}
The dynamical structure factor \cite{MullerDynamical} is defined as 
\begin{equation}
S^{\alpha\alpha}(q,\omega,M)=\sum_{n}\frac{\mid\bra{\psi_{n}} S_{q}^{\alpha} \ket{\psi_{0}}\mid^{2}}
{E_{n}-(E_{0}+\omega)+i\eta},
\label{eq7}
\end{equation}
\noindent
where $\ket{\psi_{0}}$ and $\ket{\psi_{n}}$ are the gs wavefunction for fixed $S^z=M$ and nth excited states for 
same $M$ or $M\pm 1$, respectively. $S_{q}^{\alpha}$ is defined as, 
$S_{q}^{\alpha}= (\sqrt{2\pi/{N}}) \sum_{j}S_{j}^{\alpha}e^{iqj}$, where $\alpha=x,y$ and z component. 
$E_0$ and $E_n$ are the gs and $n^{th}$ excited state energies, respectively, $\omega$ is the energy transferred 
to the spin lattice. $\eta$ is broadening factor and is fixed at 0.1 for all the calculations.\\
\begin{figure}[t]
\begin{center}
\includegraphics[width=0.5\textwidth]{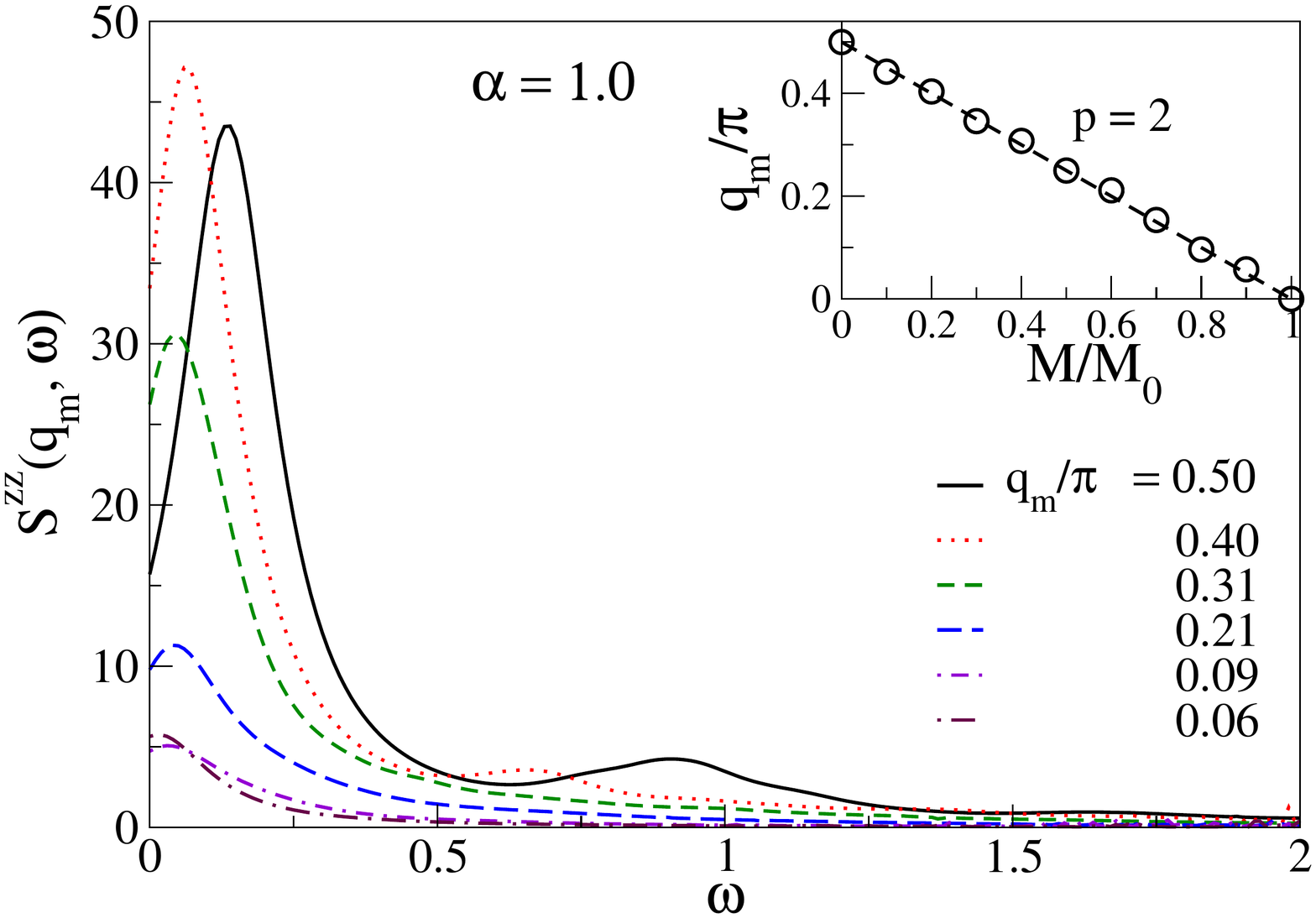}
\caption{The longitudinal dynamical structure factor $S^{zz}(q_m,\omega)$ for $\alpha=1.0$ as a function of 
$\omega$ with PBC using dynamical DMRG method is shown in the main figure for finite system size $N=104$. 
The maximum value of $q_{m}/\pi$ is calculated from $S^{zz}(q,\omega)$ and $q_{m}/\pi=0.50,0.40,0.31,0.21,0.09$ 
and 0.06 for $M=0.0,0.1,0.2,0.3,0.4$ and 0.45, respectively. Inset shows $q_{m}/\pi$ as a function of $M/M_{0}$ 
for $\alpha=1.0$ and dotted line is fitted line with $\frac{q_{m}}{\pi} = \frac{1}{p}(1-\frac{M}{M_0})$ where $p=2$.}
\label{fig7}
\end{center}
\end{figure}
\indent
The dynamical structure factor $S^{zz}(q_m,\omega)$ is shown in Fig. \ref{fig7} for 
$\alpha=1.0$. The $S^{zz}(q_m,\omega)$ for the system of size $N=104$ with given $M$ represents 
structure factor for a given value of momentum $q_m$ for which the intensity is the highest. The 
$S^{zz}(q_m,\omega)$ for $q_m$ is shown as a function of $\omega$ for different 
$M=0.0,0.1,0.2,0.3,0.4$ and 0.45. As $M$ increases the peak position of $S^{zz}(q_m,\omega)$ shifts 
towards lower $q_{m}$ and $\omega_{m}$. However, the longitudinal spin excitation is gapless in the SN/SDW$_2$ 
in the thermodynamic limit. For $M=0.0$, $q_{m}/\pi$ is at 0.5 and $q_{m}$ decreases with increasing $M$. 
In the inset of Fig. \ref{fig7} open circle represents the $q_{m}$ for different values of $M$. 
The calculated $q_m$ is fitted with a function $q_{m}/\pi = (1-{M}/{M_0})/2$. These features 
of SN/SDW$_2$ phase is directly examined by inelastic neutron scattering experiment in the presence of 
magnetic field $h$.\\
\indent
The existence of SN phase in a real material like LiCuVO$_4$ is confirmed in the presence of high magnetic field $h$. 
This material consists of planar arrays of spin-1/2 copper chains with a ferromagnetic nearest neighbor $J_1$ and 
antiferromagnetic next nearest neighbor exchange interactions $J_2$. The exchange interaction strengths are 
$J_1=-1.6$ meV and $J_2=3.8$ meV, found by fitting the data of INS and other experiments \cite{EnderleLiCuVO4}. \\
\indent
We use these parameter values for our calculations. The dynamical structure factor 
$S^{zz}(q,\omega)$ in the absence of the magnetic field is shown in Fig. \ref{fig8}. The intensity is shown by 
the contour plots. The experimentally observed $S(q,\omega)$ in figure 2 of ref. \cite{EnderleDynamic} shows as a 
function of $q$ and $\omega >3$ meV. The random phase approximation (RPA) calculation shows continuous intensity 
below $\omega <5$ meV, whereas experimental data shows high intensity between $\omega=3$ to 5 meV with momentum 
between $q/\pi=0.2$ and 0.5. The experimental data is restricted to $\omega \geq 3$ meV and shows only higher level 
of excitations. For better resolution of intensity we plot the logarithm of $S(q,\omega)$ intensity in Fig. \ref{fig8}. 
Our DMRG calculations shows that the most intense peak is at $q_m=\pi/2$ and $\omega=0.3$ meV. In fact, there are 
several values of $q$ and $\omega <3$ meV at which this system shows significant intensity of $S^{zz}(q,\omega)$. 
The $S(q,\omega)$ follows the the sum rule and we notice that major part of the intensity sum is limited to smaller 
$\omega$, and intensities of $S(q,\omega)$ observed experimentally are only a small fraction of the total intensity. 
Actually, this is easily justified by the slow variation of intensity for $\omega >3$ meV in experimentally observed 
$S(q,\omega)$. \\
\begin{figure}[t]
\begin{center}
\includegraphics[width=0.5\textwidth]{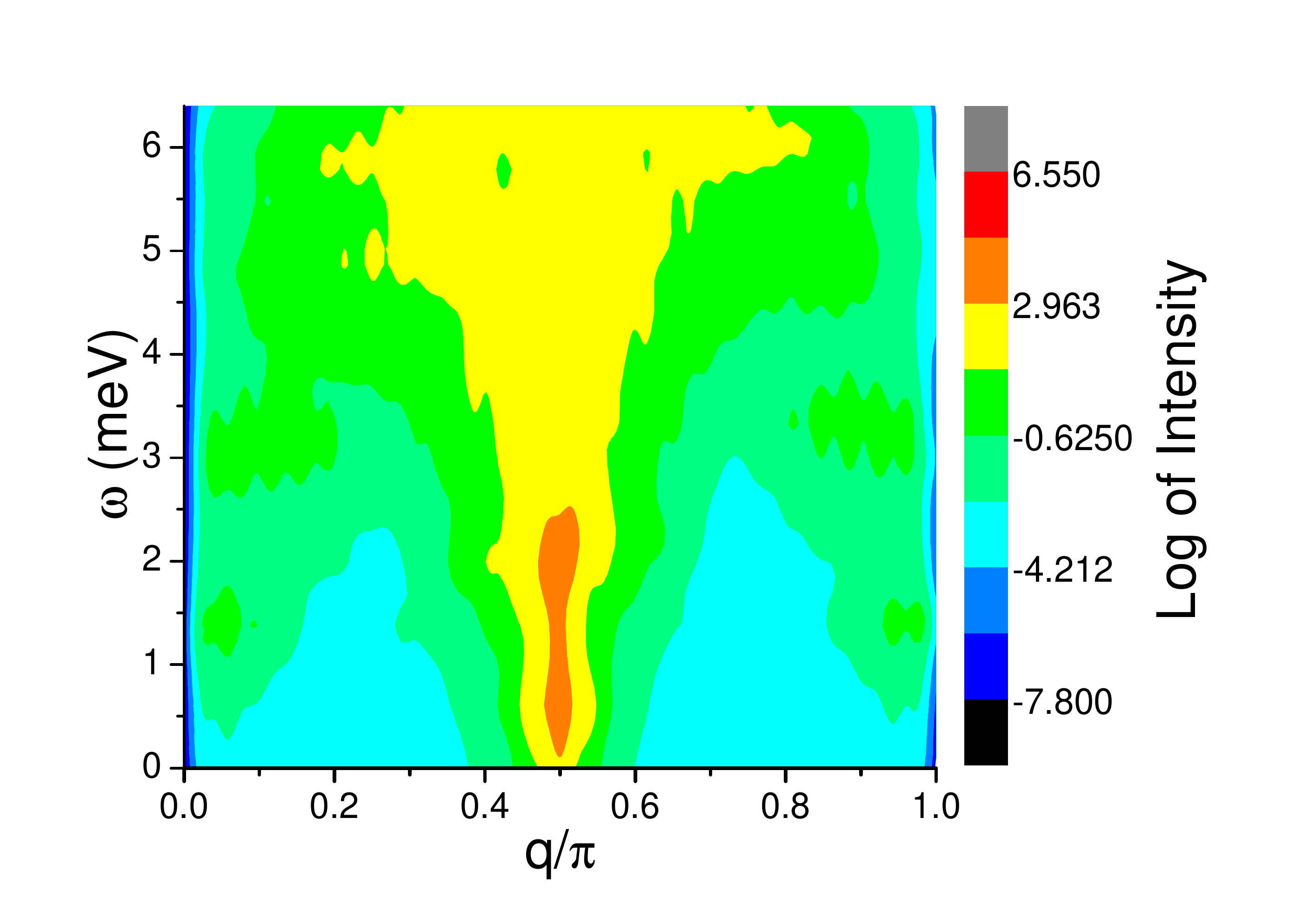}
\caption{Calculated Longitudinal dynamical structure factor of LiCuVO$_4$ as a function of wave vector 
$(q/\pi)$ and energy $(\omega)$. The color box is the longitudinal dynamical structure factor. The logarithm 
of $S(q,\omega)$ is shown for better resolution of intensities. The calculations are done for $J_{1}= -1.6$ meV, 
$J_{2}=3.8$ meV for $M=0.0$.}
\label{fig8}
\end{center}
\end{figure}
\indent
The binding energy $|E_b|$ and momentum $q_m$ in the presence of magnetic field are two important quantities 
to characterize the SN phase. The INS experiment on LiCuVO$_4$ by Mourigal {\it{et al.}} in ref. 
\cite{MourigalFieldMom} shows the linear variation of momentum $q$ with magnetic field in high magnetic field 
$h$ limit. However, $q$ is independent of field below $h=8 T$. Our results for the $|E_b|$ and momentum $q_{m}$ 
are shown in Fig. \ref{fig9} (a) and (b) for two system sizes $N=104$ and 168. The $q_m$ for LiCuVO$_4$ as a 
function of magnetization is shown in Fig. \ref{fig9} (b) for $T$=0 K. We notice that $q_{m}$ follows the linear 
relation with $M$ with a slope of $1/p=0.5$. The linear dependence of momentum is followed in the full range of $M$. 
The $|E_b|$ is shown in Fig. \ref{fig9} (a) as a function of $M/M_{0}$. For $M=0$, $|E_b|$ vanishes and it increases 
with $M$ up to $M \approx 0.4$ and then remains constant and it increases thereafter. 
\begin{figure}[t]
\begin{center}
\includegraphics[width=0.5\textwidth]{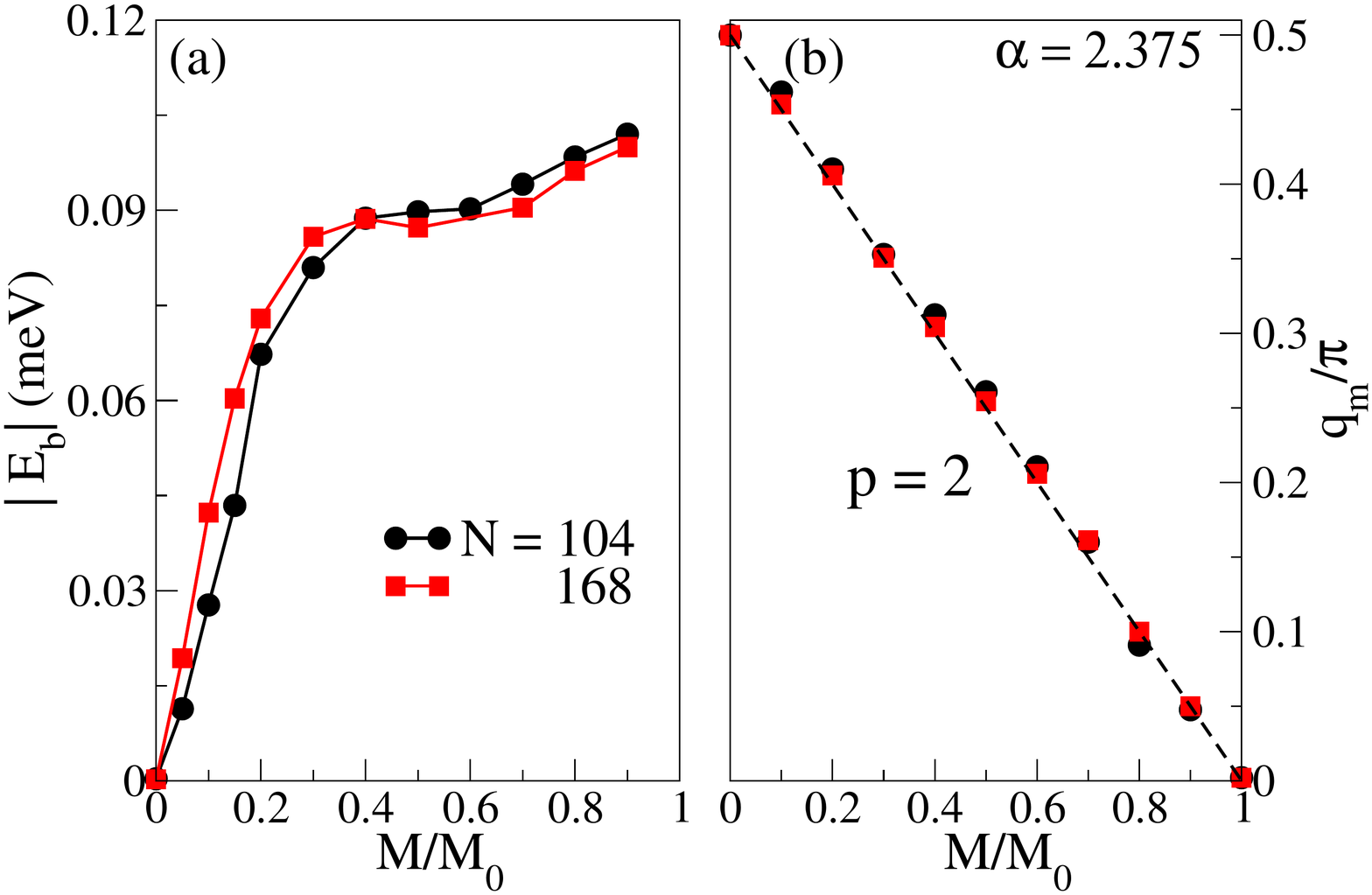}
\caption{(a) The binding energy $|E_b|$ of LiCuVO$_4$ sample with $J_1=-1.6$ meV, $J_2=3.8$ meV as a 
function of magnetization for the system sizes $N=104$ and 168 using DMRG with OBC is shown. (b) The 
momentum $q_m$ as a function of magnetization $M/M_0$ for the system sizes $N=104$ and 168 using 
DMRG with PBC is shown. The dotted line is fitted by $\frac{q_m}{\pi} = \frac{1}{p}(1-\frac{M}{M_0})$ 
where $p=2$.}
\label{fig9}
\end{center}
\end{figure}
\subsection{Dimers in spin nematic phase}
\label{Sec:C}
In the paper by Chubukov, he suggested the existence of dimerized uniaxial SN phase which is different 
from the conventional dimerization where the two nearest spin form singlet pair \cite{Chubukov1991}. In this type 
of dimerization state two neighboring spin forms spin $S=1$ state. The gs wave function is written as 
\begin{equation}
\begin{aligned}
\ket{\psi_{gs}} & = \prod_{n=2L}\{n,n\pm 1\},\{i,j\} \\
& = \prod_{n=2L}({\ket{1}_{n,n\pm 1}}+\eta{\ket{-1}_{n,n\pm 1}})/\sqrt{1+\eta^{2}},
\label{eq8}
\end{aligned}
\end{equation}
where $\ket{1}$ and $\ket{-1}$ are $\ket{\uparrow \uparrow}$ and $\ket{\downarrow \downarrow}$ triplet 
states, respectively. Although, bosonization calculation by Hikihara {\it{et al.}} suggests that 
dimerization is proportional to $\cos(a{\phi_{-}}+\pi M)$ and their average vanishes to zero \cite{HikiharaMP2008}. \\
\indent
We notice that gs is doubly degenerate in odd $S^z$ in a finite system with PBC for $\alpha>0.5$ \cite{AslamJMMM}. 
These two degenerate gs have opposite inversion symmetry. We notice that these degeneracies are independent of 
system size. In large $J_2$ limit this system is mapped to a zigzag chain with leg A and B. In the odd $S^z$ sectors 
the difference between the total spin densities on each leg A and B differ by 1. Therefore, the extra magnon 
is confined to either leg A or B depending on the symmetry of the system \cite{AslamJMMM}. Now the broken symmetry 
state is defined as $\ket{\psi_{\pm}}=\frac{1}{\sqrt{2}}(\ket{\phi_{+}} \pm \ket{\phi_{-}})$, where 
$\ket{\phi_{+}}$ and $\ket{\phi_{-}}$ are degenerate states with $+$ and $-$ inversion symmetry. Dimer order parameter 
$B_{pbc}$ for periodic system is defined \cite{ManuModfDMRG} as
\begin{equation}
B_{pbc}=\bra{\psi_{+}}(S_{i} \cdot S_{i+1}-S_{i+1} \cdot S_{i+2})\ket{\psi_{-}}.
\label{eq9}
\end{equation}
\begin{table}[t]
\centering
\caption{$B_{pbc}$ for three values of $\alpha$ are shown for $N=24$ and 28. Eq. \ref{eq9} is used to calculate 
the $B_{pbc}$ in various $S^z$ sectors.}
\begin{tabular}{|>{\centering\arraybackslash} m{1.0cm}|>{\centering\arraybackslash} m{1.0cm}|>{\centering\arraybackslash} 
m{1.8cm}|>{\centering\arraybackslash} m{1.8cm}|>{\centering\arraybackslash} m{1.8cm}|}
\hline
\multirow{2}{*}{$N$} & \multirow{2}{*}{$S^z$} & \multicolumn{3}{c|}{$B_{pbc}$} \\ \cline{3-5} 
& & $\alpha =0.8 $ & $\alpha = 1.0$ & $\alpha = 3.0$ \\ [0.5ex]
\hline
24 & 1.00 & 0.10428 & 0.10820 & 0.12391 \\
28 & 1.00 & 0.08997 & 0.09343 & 0.10763 \\ \hline
24 & 3.00 & 0.10596 & 0.11176 & 0.13055 \\
28 & 3.00 & 0.09158 & 0.09672 & 0.11390 \\ \hline
24 & 5.00 & 0.09989 & 0.10933 & 0.13283 \\
28 & 5.00 & 0.08831 & 0.09633 & 0.11786 \\ \hline
24 & 7.00 &   -     & 0.09673 & 0.12705 \\
28 & 7.00 &   -     & 0.08963 & 0.11706 \\ \hline
\end{tabular}
\label{tb2}
\end{table}
\indent
The values of $B_{pbc}$ along the rung for odd $S^z$ sectors defined in Eq. (\ref{eq9}) for PBC systems are listed in 
table \ref{tb2} for $N=24$ and 28. We notice that $B_{pbc}$ is approximately constant $0.1 \pm 0.02$ with $M$ and 
decreases with $N$ as shown in the table \ref{tb2}. All the values of $B_{pbc}$ along the leg are zero.\\
\indent
We have also calculated the dimer order parameter $B$ in OBC system in even $S^z$ sector. In this sector gs is non 
degenerate and show spiral behavior. We have followed standard procedure to calculate $B$ in 
ref. \cite{WhiteAffleck,ManuScaling2007}. The $B$ shows non-monotonic behavior with system size and it is small 
for large system. 
\section{Quadrupolar phase in spin-1}
\label{Sec:IV}
\begin{figure}[t]
\begin{center}
\includegraphics[width=0.5\textwidth]{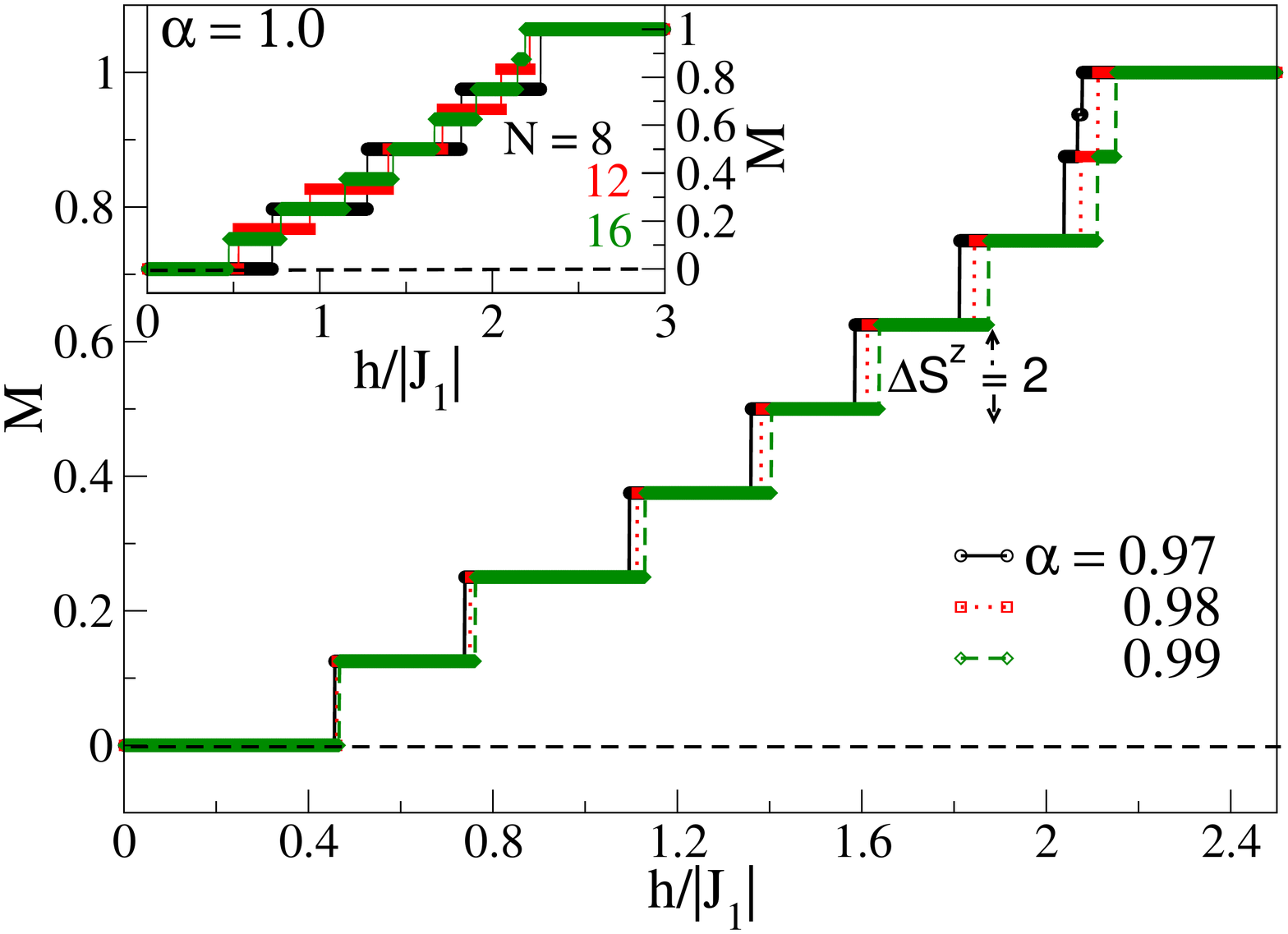}
\caption{The $M-h$ curve is shown in the main figure for $\alpha=0.97, 0.98$ and 0.99 and $N=16$ chain with PBC using 
the ED for spin $S=1$. In the inset the 
finite size effect of $M-h$ plot is shown for $N=8,12$ and 16 at $\alpha=1.0$.} 
\label{fig10}
\end{center}
\end{figure}

In this section, we explore the SN/SDW$_2$ or quadrupolar phase for spin $S=1$ for finite 
system size with PBC, and assuming the spins interaction follows the model Hamiltonian in Eq. (\ref{eq1}).\\
\indent
For the model in Eq. \ref{eq1} the ferromagnetic to singlet crossover occurs at $\alpha=0.25$ and the singlet state 
extends for all values of $J_2 >0.25$. The singlet and the triplet excitation or spin gap near the critical point 
$\alpha\approx 0.25$ is small compare to the spin gap in anti-ferromagnetic $J_1$ model. However, the 
double Haldane gap is observed in large $\alpha$ limit. The multipolar phase of higher order $p>2$ is observed for 
$\alpha <0.5$ which is consistent with earlier studies. The gs have spiral arrangement of the spins for $\alpha >0.25$. 
A detailed study of these properties of the system will be presented somewhere else \cite{Spin1New}. In this section SN or 
SDW$_2$ phase is explored for spin-1 chain with PBC in the large $\alpha$ limit. We notice that the energy convergence 
in DMRG calculation depends on the number of relevant degrees of freedom $m$ kept in the calculation, 
and energy of odd and even $S^z$ sectors follow the linear relation with $m$ but with different 
slopes. Therefore, we limit our calculations only to ED upto $N=16$.\\ 
\indent
The $M-h$ plot for three $\alpha=0.97,0.98$ and 0.99 for $N=16$ is shown in Fig. \ref{fig10}. The magnetic 
steps $\Delta S^z$ in chain with OBC is one, however, in PBC chain it is two. This may be because of 
the edge modes at the end of the chain in OBC case. We notice that there are elementary steps of $\Delta S^z=2$ 
in the magnetization with the magnetic field. The transition of steps $\Delta S^z=1$ to $\Delta S^z=2$ occurs 
at high magnetic field, and as $\alpha$ value increases the crossover point shifts to higher magnetic field. 
We also notice that for $\alpha >1$, all the elementary steps are $\Delta S^z=2$. In the main Fig. \ref{fig10}, 
variation of $M$ with $h$ is shown. The transition from mixed steps of $\Delta S^z=1$ and $\Delta S^z=2 $ to 
purely $\Delta S^z=2$ step occurs at $ \alpha=0.98$ for $N=16$ for different $M$ as shown in the main 
Fig. \ref{fig10}. To see the finite size effect, $M-h$ curve is plotted for $N=8,12$ and 16 for $\alpha=1.0$. 
We notice that the magnetic gaps decrease with $N$. 
\section{Discussion}
\label{Sec:V}
In this paper frustrated $J_1-J_2$ model Hamiltonian in Eq. \ref{eq1} for spin-1/2 and 1 chains is studied. Our studies 
are focused on the model with ferromagnetic NN and antiferromagnetic NNN interactions in the presence of a magnetic 
field $h$. We use the ED and the DMRG numerical techniques to solve the Hamiltonian in Eq. \ref{eq1}. Here we have 
discussed multipolar phases, and especially, focused on SN phase of this model. The pitch angle $\theta$, the binding 
energy $E_b$, the order parameter $\rho_q$ and the steps in the magnetization are used to characterize the SN phase. 
We modelled the dynamical structure factor $S(q,\omega)$ of the LiCuVO$_4$ compound using the parameter values 
$J_1=-1.6$ meV and $J_2=3.8$ meV in the literature \cite{EnderleLiCuVO4}. The quadrupolar phase in the spin-1 chain 
is also discussed in the large $\alpha$ limit. \\
\indent
The multipolar phase is characterized based on the pitch angle $\theta$ calculated from spin density and correlation 
function. We show that spin density and longitudinal spin-spin correlations are commensurate with each other as shown 
in Fig. \ref{fig1}. The pitch angle $\theta$ vs. magnetization $M$ plot shows multipolar phase of order 
up to $p=5$ at $\alpha=0.265$, however, the previous calculations by Sudan {\it{et al.}} are restricted to $p=4$ 
and all calculations were limited to system size up to $N=28$ \cite{SudanLauchili}. In this paper the DMRG calculations 
are done for system size up to $N=368$, especially in the large magnetization limit. We notice that in $M \rightarrow 0$ 
limit $\theta$ is weakly dependent on $M$, although, in the large magnetization limit, the pitch angle $\theta$ shows a 
linear behavior in the multipolar phase for $\alpha<0.60$. This result is consistent with the calculations of 
Sudan {\it{et al.}} \cite{SudanLauchili}. The junction of flat regime and linear variation of pitch angle $\theta$ is 
good estimate of the VC and the multipolar phase boundary. The variation of $\theta_T$ calculated from transverse 
correlation is almost independent of magnetization, and it is explained in terms of the finite gap and exponentially 
decaying correlation function \cite{HikiharaMP2008}. \\
\indent
The characterization of multipolar phase of order $p>5$ with approximate numerical technique is a difficult task 
because of the presence of large number of nearly degenerate states, and in this case it is difficult to get pure 
gs without using symmetry. To avoid the accuracy problem the ED is used to calculate step in $M-h$ curve. After careful 
investigation of gaps we show that the multipolar phase of order $p=12$ at $\alpha=0.256$, and $p=N/2$ for 
$\alpha< 0.254$ for $N \geq 24$. Although some of the previous works show that these are metamagnetic phases 
\cite{HikiharaMP2008,SudanLauchili}, we find these are actually higher order multipolar phases with small binding energy.\\
\indent
The binding energy $|E_b|$ in SN/SDW$_2$ phase rapidly increases with $\alpha$ initially, and it has maxima at 
$\alpha_{m}(M)$. In the large $\alpha$ limit the bond energy contribution of the rung decreases with $\alpha$, 
therefore $|E_b|$ decreases with $\alpha$. The value of $\alpha_{m}(M)$ increases with $M$, and the $|E_b|$ have a broad 
maxima as a function of $\alpha$ as shown in the Fig. \ref{fig5}. The bond energy analysis is done in Table \ref{tb1}. 
For lower $M$, transverse bond energy for legs and rungs both have contribution to $E_b$, whereas longitudinal 
contribution of rung plays major role in binding of two magnons. The contributions of legs and rungs for higher $M$ 
have similar trend except that the magnitude of longitudinal contribution decreases in leg, and it increases in rung. 
The $E^{T,L}_b$ decreases for higher $M$, whereas $E^{T,R}_b$ increases significantly. The $E^{T,R}_b$ actually weakens 
the binding of the magnons and longitudinal component try to enhance the $E_b$. The values of $E_b$ for 
$\alpha=0.5,1,1.5$ and 2 have similar value to the previous calculation by Onishi \cite{OnishiJapan}.\\
\indent 
The earlier studies of $J_1-J_2$ model of general $S$ chain show the absence of spin nematic 
phase in $S=1$ chain \cite{Arlego2011,Kolezhuk2012}. However, the study of Balents {\it{et al.}} shows 
presence of nematic phase in general spin using the Lifshitz nonlinear sigma model \cite{OlegBalents2016}. 
Our finite size calculations at $\alpha > 0.98$ show steps of 2 in $M-h$ curve and this results can be 
understood using their model \cite{OlegBalents2016}. The double Haldane phase agree with 
ref. \cite{Arlego2011,Kolezhuk2012}. However, we note doubly degenerate gs in odd $S^z$ sectors.\\
\indent
To characterize the SN/SDW$_2$ or quadrupolar phase the dynamical structure factor $S(q,\omega)$ is analyzed, and we 
notice that the momentum $q_m$ of most intense peak of $S(q,\omega)$ for a given $M$ varies linearly with $M$. This 
result can be directly confirmed by the INS experiments. The LiCuVO$_4$ is the most studied material for SN/SDW$_2$ phase, 
and we calculate the $S(q,\omega)$ in the absence of magnetic field $h$. The high energy peak is consistent with the 
earlier results, but the most intense peak is at $q=\pi/2$ and $\omega=0.4$ meV. We also predict the dependence of 
$q_m$ as a function of $M$ and the $M-h$ curve for single crystal of this compound. The linear variation of $q$ 
with magnetic field $h$ is shown for LiCuVO$_4$ at high magnetic field by Mourigal {\it{et al.}} \cite{MourigalFieldMom}. 
However, a more accurate measurement should be performed at low field to verify the theoretical predictions. For 
this material our calculation shows the linear variation of $q_m$ with $M$ for longitudinal $S^{zz}(q,\omega)$
for the given parameter in ref. \cite{EnderleLiCuVO4}. \\
\indent
In this paper the order parameter $\rho_q$ is also calculated for finite $N$. We notice that the extrapolation of 
$\rho_q$ goes to zero for $M \rightarrow 0$, but it is finite at high value of $M$. This result seems to partially 
agree with calculation of the exponent $\eta$ with $M$ \cite{HikiharaMP2008}. We find difficulties in calculating 
$\eta$ from spin density and spin correlations. Our order parameter calculation shows that existence of SN phase much 
below the saturation magnetic field. The bond dimerization in the SN phase at finite $h$ has been under debate. 
Chubukov claims that there are $S=1$ dimerization and the doubly degenerate gs \cite{Chubukov1991}, 
but the analytical calculation by Hikihara {\it{et al.}} \cite{HikiharaMP2008} shows the absence of dimerization. 
We show that the even $S^z$ have non degenerate and spiral gs. The odd $S^z$ have doubly degenerate gs for PBC system.\\ 
\indent
In conclusion, we have studied the $J_1-J_2$ model in an axial magnetic field $h$ with ferromagnetic $J_1$. The 
multipolar phases with multipole up to $p=14$ are calculated. We have analyzed $E_b$ in SN/SDW$_2$ phase, and we 
show that longitudinal energies of rung have major contribution to the $E_b$. We have shown the characterization 
of the SN phase with INS experiment and also predicted the $q_m-M$ relation. We think that the most of intensities 
of $S(q,\omega)$ in LiCuVO$_4$ is below $3$ meV and most intense peak is at $q=\pi/2$ and $\omega=0.4$ meV. In this 
paper we have shown that magnitude of dimerization is vanishingly small, and gs is doubly degenerate in the SN phase.\\
\indent
The model Hamiltonian in Eq. \ref{eq1} supports many quantum phases in 1D system. There are many open questions to 
be answered, like how to characterize the SN phase and other multipolar phases, what happens to magnon pairing in 
large $\alpha$ limit, and how to increase the binding energies of magnon pairing. The RIXS is a good experimental 
tool which may distinguish the SDW$_2$ and SN phase. In the large $\alpha$ limit the $J_1-J_2$ chain should behave 
like two decoupled chains and their behavior looks like anti-ferromagnetic Heisenberg chains. We can ask question 
like what happens to magnon as low lying excitations should be similar to Heisenberg spin-1/2 one dimensional chain.

\section*{Acknowledgement} We thank Prof. Z. G. Soos and Prof. G. Baskaran for their valuable comments. M. K thanks 
DST for Ramanujan fellowship and computation facility provided under the DST project SNB/MK/14-15/137.

\end{document}